# The redshift evolution of extragalactic magnetic fields

V. P. Pomakov[1,2]★, S. P. O'Sullivan[2]★, M. Brüggen,[3] F. Vazza,[3,4,5] E. Carretti,[5] G. H. Heald,[6] C. Horellou,[7] T. Shimwell,[8,9] A. Shulevski[9] and T. Vernstrom[6]

[1]*RWTH Aachen University, Templergraben 55, D-52062 Aachen, Germany*
[2]*School of Physical Sciences and Centre for Astrophysics & Relativity, Dublin City University, Glasnevin D09 W6Y4, Ireland*
[3]*University of Hamburg, Gojenbergsweg 112, D-21029 Hamburg, Germany*
[4]*Dipartimento di Fisica e Astronomia, Universitá di Bologna, Via Gobetti 93/2, I-40129 Bologna, Italy*
[5]*INAF Istituto di Radioastronomia, Via Gobetti 101, I-40129 Bologna, Italy*
[6]*CSIRO, Space and Astronomy, PO Box 1130, Bentley WA 6102, Australia*
[7]*Department of Space, Earth and Environment, Onsala Space Observatory, Chalmers University of Technology, SE-43992 Onsala, Sweden*
[8]*ASTRON, the Netherlands Institute for Radio Astronomy, Postbus 2, NL-7990 AA Dwingeloo, the Netherlands*
[9]*Leiden Observatory, Leiden University, PO Box 9513, NL-2300 RA Leiden, the Netherlands*

**ABSTRACT**
Faraday rotation studies of distant radio sources can constrain the evolution and the origin of cosmic magnetism. We use data from the LOFAR Two-Metre Sky Survey: Data Release 2 (LoTSS DR2) to study the dependence of the Faraday rotation measure (RM) on redshift. By focusing on radio sources that are close in terms of their projection on the sky, but physically unrelated ('random pairs'), we measure the RM difference, $\Delta$RM, between the two sources. Thus, we isolate the extragalactic contribution to $\Delta$RM from other contributions. We present a statistical analysis of the resulting sample of random pairs and find a median absolute RM difference $|\Delta\mathrm{RM}| = (1.79 \pm 0.09)$ rad m$^{-2}$, with $|\Delta\mathrm{RM}|$ uncorrelated both with respect to the redshift difference of the pair and the redshift of the nearer source, and a median excess of random pairs over physical pairs of $(1.65 \pm 0.10)$ rad m$^{-2}$. We seek to reproduce this result with Monte Carlo simulations assuming a non-vanishing seed cosmological magnetic field and a redshift evolution of the comoving magnetic field strength that varies as $(1 + z)^{-\gamma}$. We find the best-fitting results $B_0 \equiv B_{\mathrm{comoving}}(z=0) \lesssim (2.0 \pm 0.2)$ nG and $\gamma \lesssim 4.5 \pm 0.2$ that we conservatively quote as upper limits due to an unmodelled but non-vanishing contribution of local environments to the RM difference. A comparison with cosmological simulations shows our results to be incompatible with primordial magnetogenesis scenarios with uniform seed fields of order nG.

## 1 INTRODUCTION

Polarization measurements of distant radio galaxies are a powerful probe of cosmic magnetism. Analysing the properties of large-scale magnetization can help understand the origin of cosmological magnetic fields with important ramifications for fundamental physics (Subramanian 2016; Planck Collaboration XIX 2016b; Vachaspati 2021; Hosking & Schekochihin 2022). In particular, the presence of primordial magnetic fields has been shown to alleviate tension on the Hubble parameter $H_0$ (Jedamzik & Pogosian 2020). Furthermore, models of axion-like particles (ALPs) predict conversions between ALPs and high-energy photons in the presence of a background magnetic field (Horns et al. 2012). Thus, a detection of such fields will help to test dark matter models (Montanino et al. 2017).

However, the field strength of large-scale magnetic fields outside of galaxies and galaxy clusters remains poorly constrained, with values potentially ranging from tens of nano-Gauss as an upper bound, to $10^{-14}$ to $10^{-16}$ G as a lower limit (e.g. Taylor, Vovk & Neronov 2011; Alves Batista & Saveliev 2021). Moreover, the redshift evolution of the field strength is also unknown. The strength, the evolution, and the morphology of the intergalactic magnetic field (IGMF) provide diagnostics of its origin. There are three broad magnetogenesis scenarios, primordial, dynamo, and astrophysical, which we aim to distinguish between through observations of the evolution of cosmic magnetic fields. In the primordial scenario, a seed field is present everywhere at early times and is modified during structure formation. In the dynamo case, only a weak seed field is present at early times, to be amplified over the course of time e.g. through dynamo amplification in galaxies, galaxy cluster environments, and possibly in cosmic web filaments. In the astrophysical scenario, the magnetic field is generated at late times in stars and galaxies, and expelled into the intergalactic medium (IGM) by galactic outflows and AGN activity.

One method to measure the properties of the magnetic field in the cosmic web relies on the direct detection of synchrotron radiation produced by relativistic electrons in the cosmic web. It was used in a recent study by Vernstrom et al. (2021), who have estimated the IGMF strength to be ∼30–60 nG, which may be indicative of non-vanishing primordial magnetic field.

★ E-mail: valentin.pomakov@rwth-aachen.de (VPP); shane.osullivan@dcu.ie (SPO)

Alternatively, one can use the Faraday rotation measure (RM) of background radio sources, which has the advantage of probing the magnetized thermal material along the entire line of sight (LoS; Vernstrom et al. 2019; O'Sullivan et al. 2020; Amaral, Vernstrom & Gaensler 2021), and not just in locations of the cosmic web where special conditions are required to accelerate particles to relativistic energies and thus emit synchrotron radiation. The RM encodes the field strength along the LoS weighted by the thermal gas density following:

$$\mathrm{RM}_{[\mathrm{rad\,m}^{-2}]} = 0.812 \int_{z_s}^{0} \frac{n_{\mathrm{e}\,[\mathrm{cm}^{-3}]}\, B_{\parallel\,[\mu\mathrm{G}]}}{(1+z)^2} \frac{\mathrm{d}l_{[\mathrm{pc}]}}{\mathrm{d}z} \mathrm{d}z \quad (1)$$

with $n_\mathrm{e}$ denoting the free electron number density and $B_\parallel$ the magnetic field component along the LoS of length $l$ through a magnetized medium. Here, $z_s$ stands for the redshift of the source, whereas the integration variable $z$ denotes the redshift of the gas increment.

Different environments can contribute to this integral quantity along the LoS, such as the interiors of the radio sources themselves as well as their immediate surroundings (Vernstrom et al. 2019). The Milky Way can also cause significant Faraday rotation (e.g. Hutschenreuter & Enßlin 2020). Here, we focus on the RM caused by the IGM.

For this reason, we group the radio sources into random and physical pairs as in Vernstrom et al. (2019) and O'Sullivan et al. (2020) and focus only on the analysis of random pairs with the main quantity of interest being the RM difference ΔRM between the two sources in a random pair. Using ΔRM of a pair instead of the Faraday RM of individual sources reduces the uncertainties by minimizing the contributions of the Milky Way and Earth's ionosphere, thus isolating the intergalactic RM as far as possible. The principal advancement over O'Sullivan et al. (2020), henceforth abbreviated as SOS20, is the addition of redshift data (which were also present in the Vernstrom et al. 2019 study). This is crucial because as one can see from cosmological simulations (e.g. Vazza et al. 2017), the comoving IGMF has, in all magnetogenesis scenarios, a notable evolution with redshift. The precision of the individual RM measurements (of order 0.1 rad m$^{-2}$ or less; Sotomayor-Beltran et al. 2013) allows a study of the dependence of the RM on redshift. Analyses of redshift dependences of the RM have already been made (e.g. Welter, Perry & Kronberg 1984; Oren & Wolfe 1995; Blasi, Burles & Olinto 1999; Goodlet & Kaiser 2005; Bernet et al. 2008; Kronberg et al. 2008; Hammond, Robishaw & Gaensler 2012; Xu & Han 2014; Lamee et al. 2016; Pshirkov, Tinyakov & Urban 2016; O'Sullivan et al. 2017; Vernstrom et al. 2018, 2019; Lan & Prochaska 2020), however, the results do not always agree. In this work, the high RM precision provided by the LoTSS DR2 RM Grid, in combination with robust host galaxy identifications and redshifts, enables new insights using an RM pairs analysis (see Carretti et al. 2022 for an analysis of the individual RM values versus redshift).

In Section 2, we describe the observational data and our methodology. The results of the quantitative analysis of the data are presented in Section 3. In Section 4, we present the Monte Carlo simulations that model the observations along with the respective findings. These are interpreted and discussed in Section 6 followed by the conclusions in Section 7. Throughout this paper (but with the exception of Section 6.1), we assume a ΛCDM cosmology with $H_0 = 67.8$ km s$^{-1}$ Mpc$^{-1}$, $\Omega_\mathrm{m} = 0.307$, and $\Omega_\Lambda = 0.693$ (Planck Collaboration XIII 2016a).

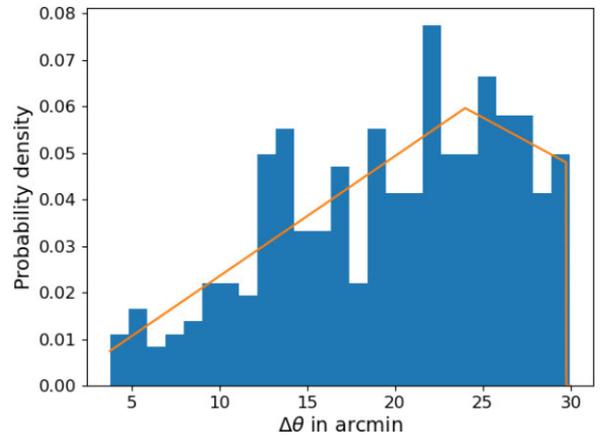

**Figure 1.** Histogram of the range of observed angular separations without distinction between RPs and PPs. This range has been divided into 25 bins of equal length. In order to draw angular separation values in the Monte Carlo simulations from this distribution (see Section 4.2), it was made continuous by fitting a normed triangular function (orange line) with a cut-off at $\Delta\theta = 29.9745$ arcmin, which is the maximum observed angular separation below our imposed upper bound of 30 arcmin.

## 2 OBSERVATIONAL DATA AND METHODS

The data used in this work come from an RM Grid catalogue (O'Sullivan et al., in preparation), derived from the LOFAR Two-metre Sky Survey Data Release 2 (abbreviated LoTSS DR2; Shimwell et al. 2022). The catalogue was constructed using RM synthesis (Brentjens & de Bruyn 2005) on the Stokes *Q* and *U* channel images of each LoTSS field, with a channel width of 97.6 kHz ranging from 120 to 168 MHz, and with an angular resolution of 20 arcsec. In total, there are 2461 RMs from extragalactic radio sources across 5720 deg$^2$ of the northern sky. All detected sources have a linear polarized intensity greater than 8 times the noise level in Stokes *Q* and *U*. We made use of the pre-existing LoTSS workflow to associate the full radio source with the polarized component(s) and to identify the host galaxies (e.g. Williams et al. 2019). This was done exclusively for the RM Grid (the full LoTSS DR2 optical/IR cross-matching effort is ongoing), in which each RM Grid source was visually classified by five different astronomers within the MKSP and SKSP teams using the LOFAR Galaxy Zoo framework. This allowed us to associate each RM Grid source with the LoTSS cataloged Stokes *I* source, and to identify the host galaxy from a combination of LEGACY optical and *WISE* infrared images (and Pan-STARRS in regions without LEGACY optical coverage). This resulted in host galaxy identifications for 2168 of the 2461 (88 per cent) of the RM Grid sources (O'Sullivan et al., in preparation), with 1046 having spectroscopic redshifts available in the literature and a further 903 photometric redshifts provided by a hybrid template fitting and machine learning approach (Duncan et al. 2021). We use all sources with redshifts in our subsequent analysis. We cross-match the RM Grid catalogue with itself to pair each polarized component with another (excluding self-matches), with the condition of the angular separation Δθ of the two components to be between 0.33 arcmin (the angular resolution of the data) and 30 arcmin (see Fig. 1 for the distribution of angular separations). The two sources can be physically related to each other (e.g. lobes or hotspots of one and the same radio galaxy), which is denoted as a 'physical pair' (PP), or they can be two entirely separate sources (e.g. lobes or hotspots of different radio galaxies that appear within the aforementioned range of angular distances from each other on the sky). The latter, called

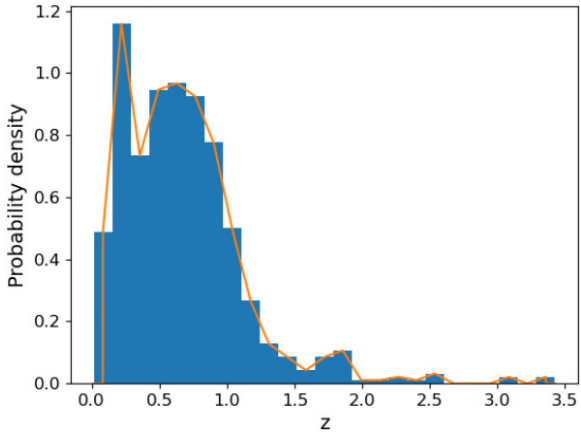

**Figure 2.** Histogram of the observed single-source redshift range divided into 25 bins of equal length. Here, no distinction has been made whether the source belongs to a physical or a random pair. In order to draw redshift values in the Monte Carlo simulations from this distribution (see Section 4.2), it was made continuous by connecting every pair of histogram values at the bin centres with a straight line (in orange).

'random pairs' (RP), is of principal interest for the present analysis because, for RPs, a strong contribution to the RM difference between the two sources $\Delta$RM is expected to stem from the IGM located between them. Since the catalogue includes sources at redshifts up to ∼3.5 (see Fig. 2 for the single-source redshift distribution), our approach is directly targeted at constraining the magnetic properties of the IGM along with their potentially redshift-dependent behaviour.

The RM pairs were classified into physical or random pairs based on their host galaxy (i.e. if each component of the RM pair has the same host galaxy it is a physical pair, otherwise it is a random pair). If at least one of the sources in a pair has no recorded redshift, the pair is discarded in order to clearly examine redshift dependences of the RM difference of random pairs. This leaves us with a final sample of 345 random pairs in which both sources have a recorded redshift and a control sample of 168 physical pairs. This is a major improvement over SOS20 who used the first data release of LoTSS and included 148 RPs without redshifts.

## 3 RESULTS FROM OBSERVATION

### 3.1 RM variation with angular separation

By imposing an upper limit of 30 arcmin on the angular separation in pairs, we ensure that the RM difference between the two sources in a pair is minimally influenced by the Milky Way's magnetic field. Of course, it is very difficult to exclude the Galactic contribution completely, as Fig. 3 shows. A rising tendency of $\Delta$RM($\Delta\theta$) is most likely caused by a non-vanishing Galactic contribution (Vernstrom et al. 2019; O'Sullivan et al. 2020). Therefore, the power-law fit on to the scatter, $k\Delta\theta^\beta$, provides an indication of the RM difference caused by the Milky Way. For the RPs with a recorded redshift for both sources, we find $k = (0.59 \pm 0.65)$ rad m$^{-2}$ arcmin$^{-\beta}$ and $\beta = 0.41 \pm 1.41$. We found that, within the errors, those parameters were indistinguishable also for the full RP sample (i.e. also including RPs without redshifts). For PPs, we find $k = (0.47 \pm 0.86)$ rad m$^{-2}$ arcmin$^{-\beta}$ and $\beta = 0.47 \pm 1.28$.

In an attempt to remove the Galactic RM (GRM) contribution, we use the latest GRM reconstruction from Hutschenreuter et al. (2022). In particular, we calculate the average GRM within a 1 deg radius surrounding each RM pair and subtract it from the corresponding RM pair value (i.e. RRM = RM − GRM). A 1 deg radius is chosen because this is the typical separation between the input data points in the Hutschenreuter et al. (2022) model (see also Carretti et al. 2022 who used a similar approach). Now, using the RRM difference, the result for RPs is $k = (0.88 \pm 0.97)$ rad m$^{-2}$ arcmin$^{-\beta}$ and $\beta = 0.20 \pm 1.41$, whereas for PPs it is $k = (0.49 \pm 0.89)$ rad m$^{-2}$ arcmin$^{-\beta}$ and $\beta = 0.44 \pm 1.28$, as illustrated by Fig. 4. We immediately note that the results for RPs are flatter after subtraction, although they remain consistent within the large uncertainties with the results before the GRM subtraction. The PPs slope is effectively unchanged. A downside of this approach is the increased uncertainties on the RRM data points due to the uncertainties in the GRM model (as illustrated by the raised dotted blue lines in Fig. 4 with respect to the ones in Fig. 3). This highlights the need for a higher areal density RM Grid for a more accurate GRM reconstruction on small angular scales. However, the total uncertainties on the individual $\Delta$RM values are still small enough to retain this approach given that it appears to have produced an improvement for the extragalactic analysis of the RP data in particular. In the following sections, while we are using the RRM values, we label them as RM for ease of notation.

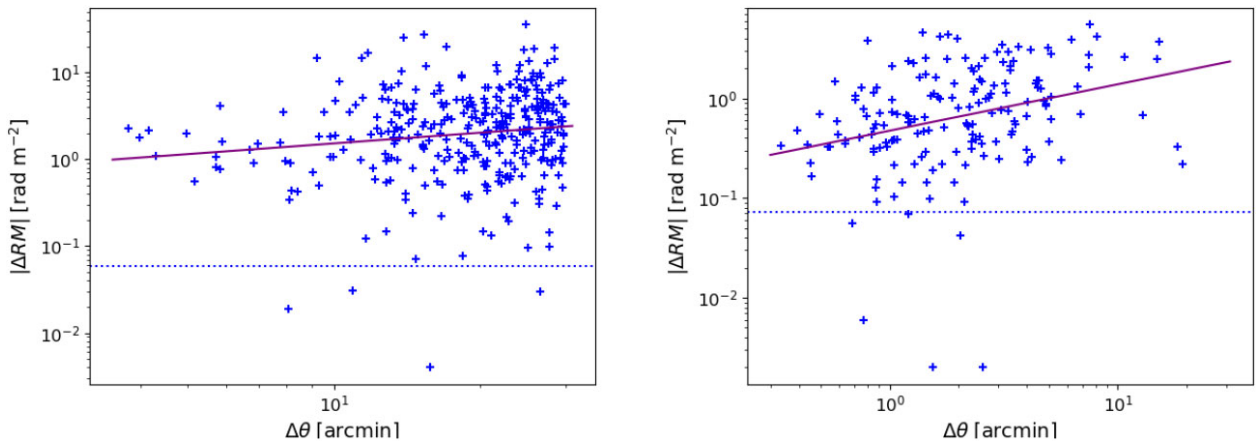

**Figure 3.** Scatter plot of the absolute RM difference in units of rad m$^{-2}$, between the two elements of random pairs (left-hand panel) and physical pairs (right-hand panel) of radio sources, versus their angular separation $\Delta\theta$ in arcmin. A power-law fit is illustrated as a solid purple line, whereas the dashed blue line represents the mean measurement error of the individual data points, which is of order 0.1 rad m$^{-2}$ or less.

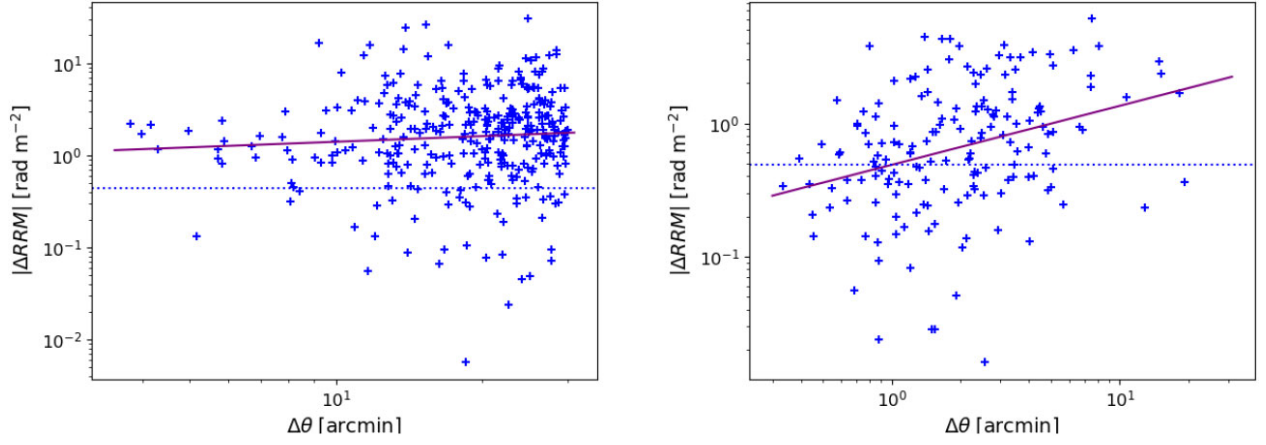

**Figure 4.** Same as Fig. 3, but after the subtraction of the GRM model of Hutschenreuter et al. (2022). The dashed blue line represents the mean measurement error of the individual data points, which is now of order 1 rad m$^{-2}$ or less.

**Table 1.** Median ('med') $|\Delta RM|_{\text{med, bin}}$ and excess median of RPs over PPs ('ex') $|\Delta RM|_{\text{ex, bin}} = (|\Delta RM|^2_{\text{med,bin}} - |\Delta RM|^2_{\text{med,PP}})^{1/2}$ of the absolute RM difference in units of (rad m$^{-2}$) for the 10 bins in $z_<$ and $\Delta z$ for RPs and in $z$ for PPs. $|\Delta RM|_{\text{med, PP}} = (0.70 \pm 0.08)$ rad m$^{-2}$. Row 1 shows the bin number, row 2 shows the $z_<$ bin centres, row 3 shows the $|\Delta RM|$ median for RPs in the respective $z_<$ bin, row 4 shows the $|\Delta RM|$ median excess of RPs in the respective $z_<$ bin over the total median of PPs, row 5 shows the $\Delta z$ bin centres, row 6 shows the $|\Delta RM|$ median for RPs in the respective $\Delta z$ bin, row 7 shows the $|\Delta RM|$ median excess of RPs in the respective $\Delta z$ bin over the total median of PPs, row 8 shows the $z$-bin centres, row 9 shows the $|\Delta RM|$ median for PPs in the respective $z$ bin.

| Bin | 1 | 2 | 3 | 4 | 5 | 6 | 7 | 8 | 9 | 10 |
|---|---|---|---|---|---|---|---|---|---|---|
| $z_<$ | 0.074 | 0.168 | 0.227 | 0.279 | 0.353 | 0.440 | 0.530 | 0.639 | 0.753 | 1.131 |
| med | 1.90 ± 0.55 | 2.00 ± 0.47 | 1.17 ± 0.17 | 1.85 ± 0.48 | 1.70 ± 0.17 | 1.59 ± 0.36 | 2.10 ± 0.41 | 2.07 ± 0.26 | 1.85 ± 0.20 | 1.26 ± 0.32 |
| ex | 1.77 ± 0.59 | 1.87 ± 0.50 | 0.93 ± 0.22 | 1.71 ± 0.52 | 1.55 ± 0.19 | 1.43 ± 0.40 | 1.98 ± 0.43 | 1.95 ± 0.27 | 1.72 ± 0.21 | 1.05 ± 0.38 |
| $\Delta z$ | 0.025 | 0.076 | 0.144 | 0.214 | 0.285 | 0.367 | 0.471 | 0.611 | 0.837 | 2.137 |
| med | 2.02 ± 0.31 | 1.28 ± 0.31 | 1.66 ± 0.28 | 1.60 ± 0.42 | 1.80 ± 0.30 | 1.66 ± 0.23 | 2.36 ± 0.52 | 1.76 ± 0.48 | 1.17 ± 0.50 | 2.15 ± 0.37 |
| ex | 1.90 ± 0.33 | 1.07 ± 0.38 | 1.51 ± 0.31 | 1.44 ± 0.47 | 1.66 ± 0.32 | 1.50 ± 0.26 | 2.25 ± 0.54 | 1.61 ± 0.52 | 0.94 ± 0.62 | 2.03 ± 0.40 |
| $z$ | 0.071 | 0.167 | 0.233 | 0.282 | 0.334 | 0.402 | 0.496 | 0.607 | 0.737 | 2.118 |
| med | 0.45 ± 0.19 | 1.05 ± 0.27 | 0.71 ± 0.17 | 0.61 ± 0.46 | 0.94 ± 0.39 | 0.70 ± 0.35 | 0.64 ± 0.09 | 0.99 ± 0.39 | 0.54 ± 0.05 | 0.72 ± 0.30 |

### 3.2 Behaviour of the absolute RM difference with redshift.

In our approach, there is a trade-off between the simplicity of the redshift-dependence analysis and isolating the IGM's contribution to $\Delta$RM. It is not trivial to analyse redshift dependences of $\Delta$RM in random pairs because there are always two sources and thus two source redshifts in each pair. We choose two separate ways to address this difficulty. Note that in both cases, the main quantity of interest is the absolute RM difference rather than the RM difference itself since the latter's expectation value along many LoSs is zero because contributions can be both positive and negative due to changes of magnetic field orientation across coherence regions.

(i) We examine the dependence of $|\Delta RM|$ of RPs on the source redshift difference $\Delta z = |z_1 - z_2|$ as follows. Apart from the computation of the entire sample's $|\Delta RM|$ statistics, we divide the sample into 10 $\Delta z$ bins such that there is a comparable amount of data points (33–36) in each bin. Then, the statistics are computed for each bin separately as well (see Table 1). A $|\Delta RM|(\Delta z)$ scatter plot of the observational data is shown in the left-hand panel of Fig. 5.

(ii) We examine the dependence of $|\Delta RM|$ on the lower of the source redshifts of the pair $z_<$ for RPs and on the sole source redshift $z$ for PPs as follows. Again, besides computing the statistics for the full sample, we furthermore bin it into 10 $z_<$ ($z$ for PPs) bins and recompute the bin statistics (see Table 1). A $|\Delta RM|(z_<)$ scatter plot of the random pair data is shown in the right-hand panel of Fig. 5 and a $|\Delta RM|(z)$ scatter plot of the physical pair data is shown in Fig. 6.

When we refer to 'statistics', unless stated otherwise, we refer to the median and an uncertainty on the median (the 68 per cent confidence interval) estimated by a bootstrapping method, rather than the root mean squared ('rms') of $\Delta$RM and its uncertainty, since the former statistics are much less sensitive to outliers (Levy 2008; O'Sullivan et al. 2020).

Table 2 presents the statistics of the absolute RM difference for the entire sample of RPs and PPs. Here, it becomes clear that the Galactic foreground correction was meaningful as it subtracted a non-negligible amount of absolute RM difference. The $|\Delta RM|$ excess of RPs over PPs is $|\Delta RM|_{\text{ex}} = (|\Delta RM|^2_{\text{med,RPs}} - |\Delta RM|^2_{\text{med,PPs}})^{1/2} = (1.65 \pm 0.10)$ rad m$^{-2}$, which remains consistent with redshift since the $|\Delta RM|$ behaviour with redshift is flat for both PPs and RPs. Keeping the uncertainties in mind, it is evident that the statistics for RPs are significantly higher than for PPs. We thus conclude that the IGM between the two sources in a random pair cannot be excluded as a significant contribution to the RM difference.

Table 3 presents the results from a Spearman rank test for correlation in $(|\Delta RM|, \Delta z)$ for RPs, $(|\Delta RM|, z_<)$ for RPs, and $(|\Delta RM|, z)$ for PPs. In all three cases, we obtain a small correlation coefficient and a $p$-value close to 1. This means that we cannot reject the null hypothesis that the observed $|\Delta RM|$ is uncorrelated with

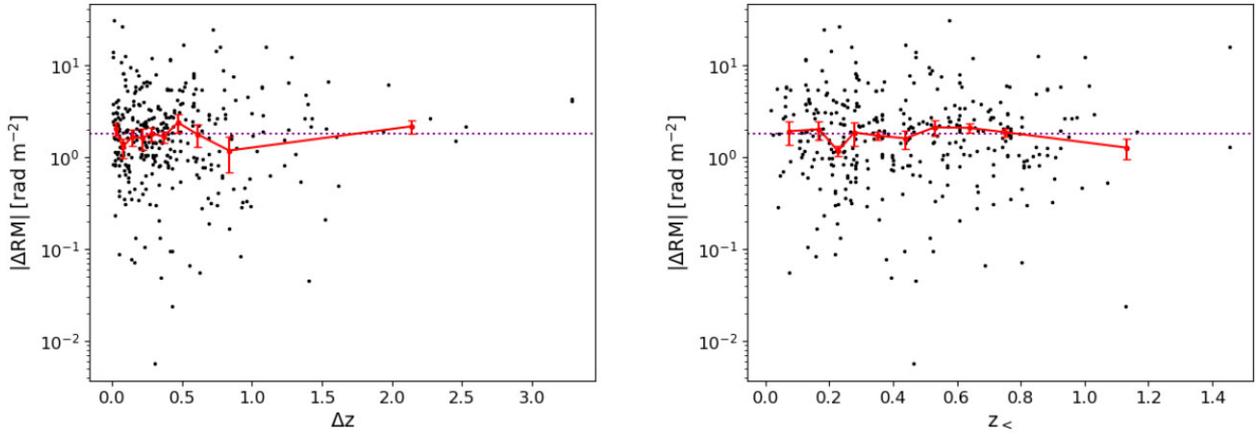

**Figure 5.** Scatter of the absolute RM difference $|\Delta RM|$ of random pairs versus redshift difference $\Delta z$ (left-hand panel) and versus the lower source redshift $z_<$ (right-hand panel) in the observational data. One black dot stands for one random pair. The red dots correspond to the $|\Delta RM|$ medians of the 10 $\Delta z$ and $z_<$ bins. Their error bars show the uncertainties estimated by bootstrapping. The dotted purple line is the median of the full random pair sample. Spearman rank tests indicate that the behaviour of $|\Delta RM|$ with respect both to $\Delta z$ and to $z_<$ is flat, as this plot also suggests.

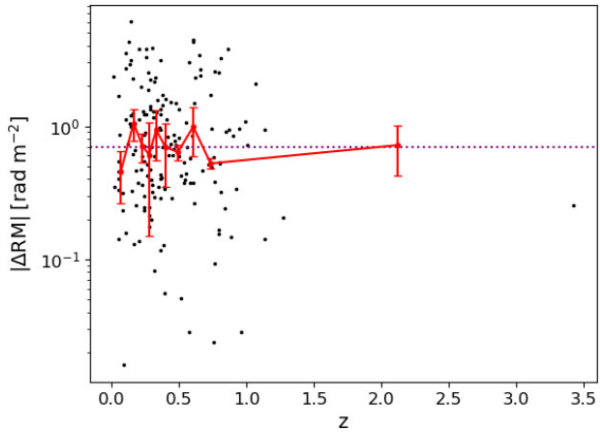

**Figure 6.** Scatter of the absolute RM difference $|\Delta RM|$ of physical pairs versus source redshift $z$ in the observational data. The representation is the same as in Fig. 5. A Spearman rank test indicates that the behaviour of $|\Delta RM|$ with $z$ is flat, as this plot also suggests.

**Table 2.** Median $|\Delta RM|_{med}$ in units of (rad m$^{-2}$) for the entire sample of RPs and PPs before and after the foreground subtraction. The uncertainty is estimated by bootstrapping.

|  | RPs before | RPs after | PPs before | PPs after |
|---|---|---|---|---|
| $|\Delta RM|_{med}$ | $2.17 \pm 0.15$ | $1.79 \pm 0.09$ | $0.68 \pm 0.06$ | $0.70 \pm 0.08$ |

$\Delta z$, $z_<$, and $z$, respectively. Essentially, this implies a flat behaviour of $|\Delta RM|$ with redshift. This result is in agreement with the single source RM analysis of NRAO VLA Sky Survey data (Hammond et al. 2012) and also with Carretti et al. (2022). Indeed, for the case of RMs generated by the IGM, an approximately flat behaviour is expected towards high redshifts, as shown by cosmological simulations (e.g. Akahori & Ryu 2011).

## 4 MONTE CARLO SIMULATIONS OF RANDOM PAIRS

Many RM analyses rely on either simple homogeneous models (e.g. Vernstrom et al. 2019) or large-scale cosmological simulations (e.g.

**Table 3.** Correlation coefficient $\rho$ and *p*-value as outputted by a Spearman rank test for correlation between: (1) $|\Delta RM|$ and $\Delta z$ for RPs, (2) $|\Delta RM|$ and $z_<$ for RPs, and (3) $|\Delta RM|$ and $z$ for PPs.

|  | (1) | (2) | (3) |
|---|---|---|---|
| $\rho$ | 0.0114 | $-0.0308$ | $-0.0496$ |
| *p*-value | 0.8333 | 0.5690 | 0.5228 |

Vazza et al. 2017). To provide a link between these two approaches and to facilitate the comparison of results with both, we opt for a computationally efficient Monte Carlo model of Faraday rotation for RPs. Our main focus is to distinguish between three broad models of magnetogenesis: (i) a strong primordial seed field (i.e. of order 1 nG), which is then amplified by structure formation, (ii) dynamo amplification of a weak seed field (e.g. $\sim 10^{-7}$ nG), and (iii) a late onset of the IGMF by injection of magnetic fields by astrophysical sources such as AGN or starburst galaxies (e.g. Vazza et al. 2017).

### 4.1 Overview of the Monte Carlo model

As in SOS20, we perform Monte Carlo simulations in a similar manner as described in Blasi et al. (1999) and Pshirkov et al. (2016). In their model, the RM difference $\Delta RM$ of a random pair is ascribed exclusively to the RM caused by the IGM between the two sources, ignoring potential contributions by the source's local environments, by the IGM between the Milky Way and the nearer of the two sources, and by the Milky Way itself. For this reason, in the context of the simulations, the RM difference $\Delta RM$ and RM will be used interchangeably. Here, we are also interested in modelling the IGM. Since we have data available on both RPs and PPs, we can best isolate the contribution of the IGM between the sources in an RP by considering the $\Delta RM$ excess of RPs over PPs $|\Delta RM|_{ex} = (|\Delta RM|^2_{med,RPs} - |\Delta RM|^2_{med,PPs})^{1/2}$ and comparing that to our simulations. In this way, we minimize the local-to-source and low-scale Milky Way effects, which are the principal contributions to the $\Delta RM$ of PPs. We do not attempt to model the local source environments and their evolution, which is beyond the scope of this work, and also beyond what cosmological simulations can feasibly provide.

The RMs of 10 000 random pairs are each computed using equation (1) by consecutively summing up the RMs induced in path-length increments of $\Delta l = \lambda_J/(1+z)$, with $\lambda_J$ denoting the redshift-dependent Jeans length $\lambda_J = 2.3/\sqrt{(1+z)}$ Mpc (Blasi et al. 1999; Pshirkov et al. 2016).

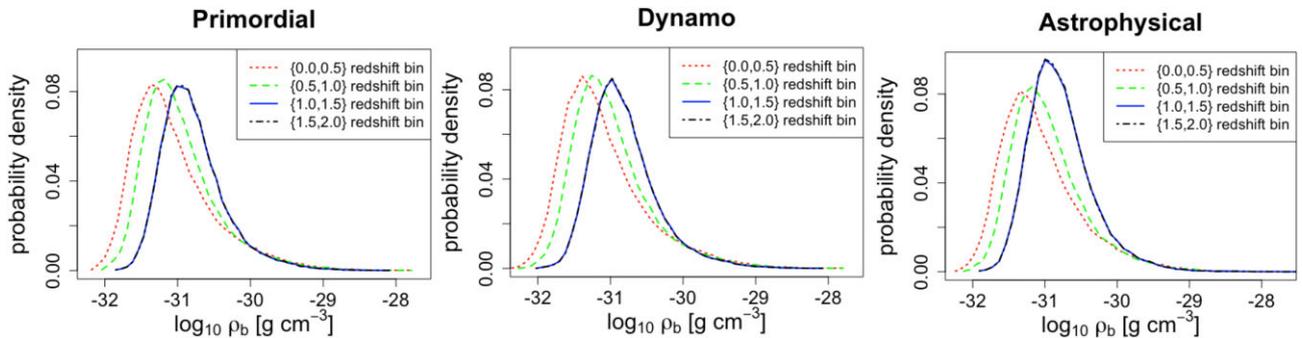

**Figure 7.** Probability density distributions for the comoving baryon mass density taken from the cosmological simulations of Vazza et al. (2017) with primordial (left-hand panel), dynamo (middle panel), and astrophysical (right-hand panel) magnetogenesis. Here, instead of the 20 redshift bins of size 0.1 that we use, for visual clarity we plot only four larger bins of size 0.5 each.

The magnetic field is tied to the electron density via $B = B_0(n_e/n_{\text{ref}})^{2/3}$ according to standard flux-freezing arguments. Here, $n_{\text{ref}}$ denotes the reference (or mean) universal number density today and is fixed at $2.0 \times 10^{-7}$ cm$^{-3}$ (e.g. Nicastro et al. 2018; Macquart et al. 2020) as opposed to $1.8 \times 10^{-7}$ cm$^{-3}$ in Blasi et al. (1999). The comoving cosmological magnetic field $B_0 = B_{\text{comoving}}(z=0)$ is a free parameter for which we explore a range of values {0.5, 1.0, 1.5, 2.0} nG in detail. After an initial exploration of lower (down to 0.1 nG) and higher (up to 10 nG) values of $B_0$, it was clear that only values around 1 nG were going to be compatible with the data, in the context of the Monte Carlo model, which is why we focus on this narrow range from here on.

The magnetic field is assumed to be coherent on coherence lengths $L_c = l_0 \Delta l$, meaning that every $L_c$ the magnetic field component projected along the LoS $B_\parallel = B \cos\phi$ receives a new, randomly generated orientation expressed by $\cos\phi$. $l_0$ is a dimensionless free parameter that we call the coherence length factor, for which we explore the values {0.1, 1, 10, 100}. For example, $l_0 = 1$ means that the magnetic field's orientation is resampled for each RM summation increment $\Delta l$, $l_0 = 0.1$ means that the magnetic field changes orientation 10 times within a summation step, and $l_0 = 10$ implies a change of field orientation only every 10 increments of $\Delta l$.

### 4.2 Model advancements in this work

The Monte Carlo simulation model has undergone multiple improvements with respect to SOS20, as is listed below.

Instead of assuming the angular separation $\Delta\theta$ for RPs to be constant at 6 arcmin, in this work it is drawn randomly from the observed distribution. For this, the observed range of angular separations was divided into 25 bins of equal size, producing the histogram in Fig. 1. In order to draw the $\Delta\theta$ values from a continuous distribution, we fit a triangular function with a cut-off to the histogram values at the bin centres and draw the values from the resulting continuous distribution.

Furthermore, instead of drawing the redshift values of radio sources from an assumed lognormal distribution, we sample them from the observed distribution made continuous in a similar fashion as for the angular separation. In the case of redshift, the distribution is not as recognizable (see Fig. 2). For this reason, instead of guessing an appropriate fit function, we take the continuous distribution to be the set of straight lines connecting two consecutive histogram values at the bin centres.

The Monte Carlo algorithm is prevented from drawing two arbitrarily close redshift values $z_1$ and $z_2$ within a random pair by imposing the lower limit $\Delta z_{\text{min}} = 0.0025$ (taken from the observational data) on their difference $\Delta z = |z_1 - z_2|$. Without this lower bound, the $\Delta$RM($\Delta z$) scatter exhibits an unrealistically large amount of points at unphysically small $\Delta z$ values with correspondingly low RM differences, skewing the distribution too far towards the low $\Delta$RM and $\Delta z$ end.

With the electron number density drawn randomly from a distribution (to be elaborated further below) and used for the scaling of the magnetic field amplitude $B = B_0(n_e/n_{\text{ref}})^{2/3}$, then $B$ can adopt a wide range of values. Physically, the magnetic field does not reach values above a given fraction of the equipartition field amplitude due to a minimum of the energy of the system of plasma and magnetic field located at this fraction (Klein & Fletcher 2015). We implement this in our Monte Carlo model by means of a fixed cut-off parameter $B_{\text{eq}} = 0.1\,\mu$G, which absorbs both the actual equipartition field amplitude and the fraction of it at which the energy minimum is located (e.g. Brown et al. 2017; Vernstrom et al. 2017; Govoni et al. 2019; Locatelli et al. 2021). If a number density is drawn which causes the field amplitude B to exceed $B_{\text{eq}}$, the algorithm sets it back to the cut-off value $B_{\text{eq}} = 0.1\,\mu$G. The values 1 and 0.01 $\mu$G were tested in preliminary simulation runs but were discovered to be too high for the field amplitude to ever reach, or reached unrealistically often, respectively. For this reason, in the final simulations, the parameter $B_{\text{eq}}$ was fixed at 0.1 $\mu$G.

#### 4.2.1 Using realistic density distributions from cosmological numerical simulations

A modification was made to the sampling of electron number density values. The reader is directed to SOS20 (O'Sullivan et al. 2020) for a detailed description of their number density drawing for the density values in each path-length increment $\Delta l$ along the two LoSs in a pair. In essence, their values were drawn from a multivariate lognormal distribution with a correlation coefficient that depends on the transverse distance $d_{\text{trans}}$ between the two LoSs at the redshift of the given increment.

Here, we draw the density values from cosmological simulations for each of the three magnetogenesis scenarios (Vazza et al. 2017), for which we have received redshift, comoving density, and magnetic field data for 100 LoSs each. Examples of the density distributions for each scenario are shown in Fig. 7 and the physics behind the differences is elaborated on in Appendix A. For the density drawing, we simply collect all density values in 20 redshift bins (explained further below) from all 100 LoSs per magnetogenesis scenario. In this manner, we are able to gather 200 000–400 000 density values for each redshift bin to draw from.

Now we must account for the correlation of the densities along the LoS for close pairs. To this end, we used the observed two-point correlation function for galaxies as a proxy, in a similar manner to SOS20, with the correlation coefficient (Mo, van den Bosch & White 2010):

$$cc = \begin{cases} \left(\frac{d_{\text{trans}}}{0.2\,\text{Mpc}}\right)^{-1.8}, & d_{\text{trans}} > 0.2\,\text{Mpc} \\ 1, & d_{\text{trans}} < 0.2\,\text{Mpc} \end{cases} \quad (2)$$

However, we could not use an analytical approach to drawing correlated samples from the density distributions derived from the cosmological simulations because they do not have a simple functional form (as opposed to the lognormal distributions in SOS20). Therefore, in order to draw correlated samples from these density distributions, we used a 'sample and iterate' algorithm to reproduce a target correlation matrix, as described in Ruscio & Kaczetow (2008). This approach allows one to sample directly from a non-normal distribution and the algorithm identifies an intermediate correlation matrix that yields the best reproduction of the target correlation matrix through an iterative, trial-and-error process.

In initial testing, we verified that the output correlated samples taken from the cosmological density distributions robustly followed any user-specified correlation coefficient. Using this code, we then drew 10 000 pairs of correlated samples from the cosmological density distributions for each magnetogenesis scenario and each redshift bin. In this way, we generated look-up tables for correlation coefficients from 0.1 to 0.9 in increments of 0.1, that could be drawn from. For each two path-length increments along the two LoS in a simulated random pair, the correlation coefficient is computed as in equation (2) and then rounded to the first decimal place, determining which look-up table the two density values of the path-length increments must be drawn from.

To account for the comoving baryon mass density's redshift evolution in accordance with structure formation, we split the simulated redshift range from Vazza et al. (2017) from 0 to 2 into 20 equal redshift bins of size 0.1 and draw each density value from the respective redshift bin depending on the redshift of the path-length increment in question. Since our redshift range reaches up to 3.5, all densities in increments with redshifts above 1.9 are drawn from the redshift bin with bounds 1.9 and 2 in the cosmological simulations. We opt against implementing an assumed extrapolation because the redshift evolution of the baryon mass density distribution becomes practically unnoticeable towards higher redshifts (see Fig. 7). The last two redshift bins in the plot exhibit a peak of the probability density of virtually the same height and position, which justifies our drawing of all densities above redshift 1.9 from the same redshift bin without having to guess an extrapolation.

Since the output of the cosmological simulations is actually the comoving baryon mass density $\rho_b$, we convert it to a comoving electron number density according to the model by Tanimura et al. (2019; see also Jaroszynski 2019), assuming standard literature values for the mean molecular weight $\mu_e = 1.14$ and thus implicitly for the primordial hydrogen and helium abundances, $X = 0.76$ and $Y = 0.24$, respectively. We note that this assumes that all data are for more recent times than the epoch of helium reionization (i.e. at $z \sim 3-4$), which is not an issue for most sources in this sample.

**Table 4.** The parameters in the Monte Carlo simulations. The relation of the first two parameters to RM is presented as in simple homogeneous models of the IGMF in accordance with O'Sullivan et al. (2020) and Vernstrom et al. (2019), whereas the effect of $\gamma$ on RM is motivated in Section 6.1. The label 'dd' corresponds to the density distributions used to draw density values from, i.e. in the *prim* case, the density distributions were taken from Vazza et al. (2017)'s cosmological simulations with a primordial magnetogenesis model, as explained in Section 4.2.1 with analogous correspondences for *dyn* and *astro*.

| Parameter | Explored values | Effect on RM |
|---|---|---|
| $B_0$ | {0.5, 1.0, 1.5, 2.0} nG | RM $\propto B_0$ |
| $l_0$ | {0.1, 1, 10, 100} | RM $\propto l_0^{1/2}$ |
| $\gamma$ | {1.5(0.5)5.0} | RM $\propto (1+z_s)^{1.5-\gamma}$ for $\gamma \neq 1.5$ |
| | | RM $\propto \ln(1+z_s)$ for $\gamma = 1.5$ |
| dd | *prim*, *dyn*, *astro* | |

After performing the comoving density draws, the density values are converted to physical densities by imposing a redshift evolution with $(1 + z)^3$. Through the scaling of the magnetic field with number density, this immediately implies redshift scaling of the physical cosmological magnetic field as $(1 + z)^2$.

### 4.2.2 Modelling the evolution of the comoving IGMF

On top of the evolution of the physical magnetic field, we allow for an additional scaling of the comoving IGMF with redshift $B_0(z) = B_0(1 + z)^{-\gamma}$ due to some still unspecified magnetic field injection, amplification or evolution mechanism. The values explored for the parameter $\gamma$ are {1.5, 2.0, 2.5, 3.0, 3.5, 4.0, 4.5, 5.0}. Values below 1.5 were not explored for theoretical reasons (see Section 6.1).

To summarize, our model consists of three free parameters: the comoving cosmological magnetic field $B_0$ at $z = 0$, the coherence length factor $l_0$, and the comoving IGMF redshift evolution exponent $\gamma$. Table 4 presents the simulated parameter ranges and, for orientation, the effect of the parameter on the RM as in simplified homogeneous models. The dependence on $B_0$ and $l_0$ has been taken from O'Sullivan et al. (2020) and Vernstrom et al. (2019)'s models. The dependence on $\gamma$ can be derived from a simple model of homogeneous comoving electron density and a flat, matter-dominated universe, yielding:

$$\text{RM} = \begin{cases} \frac{0.812}{1.5-\gamma} \frac{c}{H_0} n_{\text{ref}} B_0 \cos\phi\,(1+z_s)^{1.5-\gamma} - C, & \gamma \neq 1.5 \\ 0.812 \frac{c}{H_0} n_{\text{ref}} B_0 \cos\phi \ln(1+z_s), & \gamma = 1.5 \end{cases} \quad (3)$$

where $H_0$ is the Hubble constant, $n_{\text{ref}}$ is said homogeneous comoving electron number density, and $\phi$ is the assumed constant angle between the LoS along which the RM is computed and the magnetic field direction. The full calculation with all model assumptions is carried out in Section 6.1.

Table 4 makes clear the degeneracy between $B_0$ and $l_0$. $\gamma$ has a complex relationship with the other two parameters. For example, a higher $\gamma$ can be compensated by a higher $B_0$ or $l_0$ value. However, $\gamma$ has a stronger effect on the RM of higher redshift sources, whereas the effect of $B_0$ and $l_0$ is not redshift-dependent.

Technically, there is a fourth 'parameter', which denotes whether the density distribution from which density values are drawn stems from a primordial, dynamo, or an astrophysical magnetogenesis cosmological simulation model. Hence, the 'values' of this parameter dubbed 'dd' in Table 4 are denoted as *prim*, *dyn*, and *astro*. Their effect on RM cannot be expressed in a simple analytical formula.

**Table 5.** Results for the top five best-fitting models D1, P1, P2, D2, and A1 out of 384 models in total. Included are the full-sample absolute RM difference median, the likelihood, the correlation coefficients, and p-values from a Spearman rank test for $|\Delta\mathrm{RM}|$ with $\Delta z$ and $z_<$, respectively.

| Model: \| $dd$ \| $B_0$/nG \| $l_0$ \| $\gamma$ \| | $\|\Delta\mathrm{RM}\|_{\mathrm{med}}$ [rad m$^{-2}$] | $\log P(d\|m)$ | $\rho_{\Delta z}$ | $p_{\Delta z}$ | $\rho_{z_<}$ | $p_{z_<}$ |
|---|---|---|---|---|---|---|
| D1: \| dyn \| 2.0 \| 0.1 \| 4.5 \| | $1.52 \pm 0.03$ | $-7.86$ | 0.012 | 0.22 | 0.11 | $\ll 10^{-6}$ |
| P1: \| prim \| 1.0 \| 10 \| 4.5 \| | $1.62 \pm 0.03$ | $-8.03$ | 0.022 | 0.03 | 0.05 | $\ll 10^{-6}$ |
| P2: \| prim \| 0.5 \| 100 \| 2.5 \| | $1.42 \pm 0.03$ | $-8.35$ | 0.0007 | 0.95 | 0.09 | $\ll 10^{-6}$ |
| D2: \| dyn \| 2.0 \| 1 \| 4.5 \| | $1.48 \pm 0.03$ | $-8.84$ | 0.021 | 0.03 | 0.10 | $\ll 10^{-6}$ |
| A1: \| astro \| 2.0 \| 10 \| 3.0 \| | $1.57 \pm 0.03$ | $-8.93$ | 0.046 | $<10^{-5}$ | 0.13 | $\ll 10^{-6}$ |

## 5 RESULTS OF COMPARISON BETWEEN SIMULATION AND OBSERVATION

### 5.1 Criteria for assessing agreement between simulation and observation

To quantify the degree of agreement between simulations and observations and select a best-fitting model, we first divide the simulated spaces ($|\Delta\mathrm{RM}|$, $\Delta z$) and ($\Delta\mathrm{RM}$, $z_<$) into the same 10 respective bins in $\Delta z$ and $z_<$ as the data. As previously explained, the medians in these simulated bins are to be compared to the bin-wise $|\Delta\mathrm{RM}|$ excess medians of RPs over PPs, i.e. from each RP bin as listed in Table 1, we subtract the total PP median of $(0.70 \pm 0.08)$ rad m$^{-2}$. Note that we cannot subtract the binned PP medians since the redshift bins of RPs and PPs are not directly comparable. The reason is that the redshift values associated with RPs and PPs are different in nature. RPs have, in our analysis, a source redshift difference $\Delta z = |z_1 - z_2|$ and the redshift value of the nearer source $z_<$, whereas PPs have the single redshift value $z$ of the source. Furthermore, the bin centres of those quantities are also different in value for RPs and PPs because the bins were chosen in such a way that each bin contains a comparable number of pairs. This is why we opt for subtracting the overall PP median from each RP bin, making use of the finding that the $|\Delta\mathrm{RM}|$ behaviour of PPs with respect to $z$ is flat.

Then, we compute the difference between the simulated and the observed bin median normed by the bin uncertainty of this median difference. The bin uncertainty is calculated as the square root of the quadratic sum of the bootstrap error of the simulated and the observed median. From this, we compute the likelihood to obtain the data $d$ given the respective model $m$ from

$$P_i(d|m) = \frac{1}{2\pi\sigma_{a_{\mathrm{tot,i}}}\sigma_{b_{\mathrm{tot,i}}}} \exp\left(-\frac{(a_{\mathrm{d,i}} - a_{\mathrm{m,i}})^2}{2\sigma_{a_{\mathrm{tot,i}}}^2} - \frac{(b_{\mathrm{d,i}} - b_{\mathrm{m,i}})^2}{2\sigma_{b_{\mathrm{tot,i}}}^2}\right), \quad (4)$$

where

(i) $a_{\mathrm{m,i}}$ is the simulated $|\Delta\mathrm{RM}|$ median in a given $\Delta z$ bin indicated by the index $i$ and $a_{\mathrm{d,i}}$ is the respective observed median.
(ii) $b_{\mathrm{m,i}}$ is the simulated $|\Delta\mathrm{RM}|$ median in a given $z_<$ bin indicated by the index $i$ and $b_{\mathrm{d,i}}$ is the respective observed median.
(iii) $\sigma_{a_{\mathrm{tot,i}}} = \sqrt{\sigma_{a_{\mathrm{m,i}}} + \sigma_{a_{\mathrm{d,i}}}}$, as was already mentioned, with an analogous expression for $\sigma_{b_{\mathrm{tot,i}}}$

In the end, to find the likelihood for a model $P(d|m)$, one must build the product of the ten $P_i(d|m)$ for the 10 bins.

The likelihood is computed for each model, i.e. for each combination of values of the four parameters as listed in Table 4. This way, the simulation run with the best agreement with the data as quantified by the maximum natural logarithm of the likelihood $\log P(d|m)$ is selected as the best-fitting model.

After the best-fitting model has been selected in this way, we perform the following check. Since we could not reject the hypothesis that the data were flat in the ($|\Delta\mathrm{RM}|$, $\Delta z$) and ($\Delta\mathrm{RM}$, $z_<$) spaces (Table 3), we perform Spearman rank tests for all simulation models with the aim of confirming that the hypothesis of flatness cannot be rejected there either.

The uncertainties on the best-fitting parameter values were estimated as follows: The simulated parameter values of $B_0$ and $\gamma$ are equally spaced at intervals of 0.5. If a simulated parameter value $a$ is singled out as a best-fitting value, the real best-fitting value has a uniform probability of being anywhere between $a - 0.25$ and $a + 0.25$. Thus, for the uncertainty on $a$ we take the standard deviation of a uniform distribution with range 0.5, i.e. $0.5/\sqrt{12} \approx 0.14 \approx 0.2$. In the final step, the result has been rounded *up* to the first decimal place in order not to starkly underestimate the uncertainties that come also from the simplifications in the model and the degeneracies of the parameters, discussed in greater detail in Section 5.3. The process of error estimation is similar for $l_0$ except that the parameter values are equidistant in decadic-logarithmic space, meaning that a uniform standard deviation can be determined for $\log_{10} l_0$, which must then be propagated on to $l_0$ assuming Gaussian error propagation.

### 5.2 Best-fitting results

Henceforth, the models will be referred to based on the corresponding values of the parameters using the shorthand notation $|\ dd\ |\ B_0/\mathrm{nG}\ |\ l_0\ |\ \gamma\ |$. The explored values of these parameters yield a total of 384 competing models.

The utilized maximum likelihood method selects $|$ dyn $|$ $2.0 \pm 0.2$ $|$ $0.10 \pm 0.07$ $|$ $4.5 \pm 0.2$ $|$ as a best-fitting model that we dub D1 (i.e. the model with the largest $\log P(d|m)$ value). The results are summarized in Table 5, which, apart from D1, includes also results for the four next-to-best-fitting models P1, P2, D2 and A1 for comparison. Fig. 8 shows the ($|\Delta\mathrm{RM}|$, $\Delta z$) and ($|\Delta\mathrm{RM}|$, $z_<$) scatters for D1 as a representative for *dyn* models. Fig. A1 shows, analogously, the results for P1 standing for *prim* models and Fig. A2 for A1 as an *astro* model.

As can be inferred from the plots and the table, our model is capable of providing an overall good fit. The best-fitting model's $|\Delta\mathrm{RM}|_{\mathrm{med}}$ is $\sim 1\sigma$ below the data's excess median of $(1.65 \pm 0.10)$ rad m$^{-2}$, which is satisfactory given that the IGM is the only contribution we consider. The plots highlight the flat behaviour of the absolute RM difference of the model versus both $\Delta z$ and $z_<$ with the notable exception of the low $z_<$ range where we observe a drop of $|\Delta\mathrm{RM}|$. We will proceed with a critical analysis of these general statements shortly.

We first report a general preference for *dyn* models as selected by our maximum likelihood criterion. In fact, there are a total of 18 *dyn*, 12 *prim*, and 12 *astro* models with likelihoods $>10^{-5}$.

Furthermore, the results exhibit a preference for the higher values of $\gamma$ of our simulation range because higher $\gamma$ values ensure a flatter

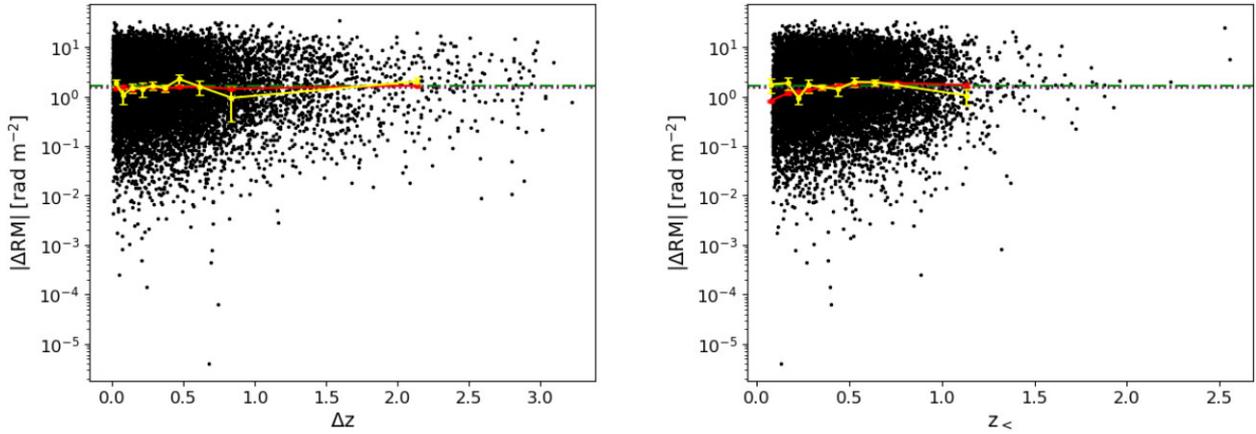

**Figure 8.** Scatter of the absolute RM difference of random pairs versus redshift difference (left-hand panel) and versus the lower source redshift (right-hand panel) for the best-fitting result | dyn | 2.0 | 0.1 | 4.5 |. One black dot stands for one simulated random pair. The red dots correspond to the |ΔRM| medians of the 10 simulated $\Delta z$ and $z_<$ bins. Their error bars show the uncertainties estimated by bootstrapping. The dotted purple line is the median of the full simulated random pair sample. To serve as comparison, the dot–dashed green line represents the observational data's full-sample excess median and the yellow dots with errorbars show the observed binned excess medians.

|ΔRM| behaviour by reducing the RM at higher $z$ more than at low $z$. At the same time, this leads to a suppression of the median |ΔRM|$_{med}$ to about $1\sigma$ away from the excess median. While this is not a statistically significant difference, it indicates that our models cannot provide an ideal fit in terms of both flatness and the median value at the same time. The maximum likelihood method seeks to balance the suppression of the median resulting from high $\gamma$ by higher values of either $B_0$ or $l_0$. The fact that such 'compensation' effects between the parameters are possible is a caveat of the simulation model that we discuss in Section 5.3 along with a plausible solution to the trade-off between flatness and the median value.

Now turning to the results of the Spearman rank tests in greater detail, we note that best-fitting model D1 has a Spearman rank correlation coefficient with respect to $\Delta z$ of $\rho_{\Delta z} = 0.012$ and a respective p-value $p_{\Delta z} = 0.22$ indicating that the null hypothesis of flatness with respect to $\Delta z$ cannot be rejected. In fact, neither of the top four best-fitting models shows a statistically significant correlation in (|ΔRM|, $\Delta z$) space, whereas A1 does, which is true for 7 out of the 12 *astro* models with likelihoods $>10^{-5}$. It is furthermore worth noting that, although the *dyn* model D1 is selected as best fit by the maximum likelihood criterion, there are *prim* models, like P2, that generally exhibit a correlation coefficient in $\Delta z$ that is 1 or even 2 orders of magnitude lower. The fact that the null hypothesis of flatness cannot be rejected for all top five models but A1 remains valid, though.

In contrast, all 384 models exhibit a statistically significant correlation in (|ΔRM|, $z_<$) space. This deviation from flat behaviour is visibly present only at low $z_<$ as we see from Figs 8, A1, and A2. D1's correlation coefficient with respect to $z_<$ is $\rho_{z_<} = 0.11$ and the p-value is $p_{z_<} \ll 10^{-6}$, indicating a low but highly statistically significant correlation between |ΔRM| and $z_<$. There exist primordial models, like P1, that perform slightly better in this respect too, exhibiting correlation coefficients of one order of magnitude lower. This shows again that our models cannot provide an ideal fit at the same time in terms of both flatness and the likelihood value.

### 5.3 Caveats of our model and methodology

Before moving on to the discussion of these results, it is important to be explicit about the caveats of our simulation model and methodological approach. The simulation model has two main issues: the complex (if not directly degenerate) relationships and compensation effects between the parameters, and the disregard of all other RM contributions except for the intergalactic contribution on ∼Mpc scales and larger.

#### 5.3.1 Model degeneracies

In general, results from a model with degenerate parameters must be considered with an openness for larger uncertainties than the ones given for the best-fitting parameter values in Section 5. Due to the degeneracy between $n_e$, $B_0$, and $l_0$, the best-fitting values for the latter two quantities depend on the choice of density drawing. For example, with the densities drawn from the cosmological simulations, lower values for $B_0$ and $l_0$ are preferred than for the lognormal density drawing of SOS20. The reason for this is the longer high-density tail of the density distributions from the cosmological simulations (see Fig. 7) than of a lognormal distribution. Furthermore, $B_0$ and $l_0$'s degeneracy (recall Table 4) also leads to compensatory effects between the two parameters: a smaller comoving magnetic field can be compensated by a larger coherence length factor and vice-versa. Most importantly, however, as we observe in the simulation results, $\gamma$ exhibits compensatory effects with the other two free parameters. For $\gamma$ these effects are redshift-dependent, though: a larger value reduces the high-$z$ RM more than the low-$z$ RM, which is a way to obtain a flatter behaviour of RM with redshift. At the same time, $\gamma$ suppresses the overall RM, so in most preferred models, higher $\gamma$ values are compensated by higher values for either $B_0$ or $l_0$. In the following, we will elaborate on how all of those compensatory effects might lead to a distortion of the results due to the fact that our simulation model considers only the RM contribution of the IGM.

#### 5.3.2 RM contributions local to the radio source

Our simulations have two main inconsistencies with the data: they exhibit a small, but statistically significant correlation with $z_<$ due to a drop of |ΔRM| at low $z_<$, and the simulated medians tend to be around $1\sigma$ away from the median in the data. This is not a statistically significant difference but it could be indicative of the need to consider a small contribution from local environments too, as modelled for example by Mahatma et al. (2021). Carretti et al. (2022) analyse that the LoTSS DR2 ΔRM coming mostly from lower density environments. This potentially rules out any high local contributions[1] and reinforces our model assumption of the IGM being the predominant contributor, however, without excluding the possibility of a small local contribution, which could be the key to bridging the gap of about $1\sigma$ between our simulations and the data.

Indeed, the analysis of the physical pairs as a control data sample showed with its non-vanishing |ΔRM| median (Table 2) that the IGM between the two sources in a random pair cannot be the sole contribution to the RM. While physical pairs also have the IGM between them and the observer as a contribution, it is a very small one due to the closeness of the two LoSs in a PP. Therefore, the physical pair result must be coming mostly from local environments and small-scale Galactic variations. By using the excess |ΔRM| of RPs over PPs when comparing simulations to data, we essentially removed some local-to-source effects and Galactic variations on small angular scales in the aim of further isolating the contribution of the IGM as best as possible in the data. However, some local contributions are bound to remain since they are not the same for RPs and for PPs because the RM variation between sources at different redshifts is likely to be different than RM variations within the same source.

A variation in the source-to-source local environments could be the solution to the simulations' drop of |ΔRM| at low $z_<$ and essentially improve our simulations' flatness. Goodlet & Kaiser (2005) find evidence for an increase in the local environment RM with redshift for their sample of 26 sources. Such a trend, in a pair analysis, would introduce an extra contribution to the RM difference at low $z_<$ coming from the fact that the nearer source (the one with redshift $z_<$) has a lower local contribution than the higher redshift source. A detailed Faraday depolarization study with matched angular resolution at higher frequencies would provide a means to better constrain the effect of the local environment. Such a study may be possible with APERTIF (van Cappellen et al. 2022) in the near future.

In summary, our simulation models generally exhibit two main problems: a deviation from flatness at low $z_<$ and an overall slight underestimation of the |ΔRM| median, be it for the full sample or for the various bins in $z_<$ and $\Delta z$. The maximum likelihood selection criterion seeks to remedy these problems by preferring models with higher $\gamma$ values to try to yield flatness, and then also with higher $B_0$ or $l_0$ values (which are degenerate) in order to compensate the overall suppression of the RM that results from higher $\gamma$. By considering a non-negligible contribution from local environments, both problems can be solved without having to resort to extremes of the parameter values. However, this would require detailed simulations of the local effects of radio galaxies embedded in a realistic distribution of environments in the cosmic web. In this sense, it might be more reasonable to quote our best-fitting results merely as upper limits on the parameter values for the IGM. Furthermore, this means we must be critical about our model's capability to discriminate between different magnetogenesis scenarios. We saw that the inclusion of local-to-source effects could significantly improve the flatness behaviour, however, it is unclear how it will affect the model selection via the maximum likelihood method and whether the preference for a given magnetogenesis scenario would change. Section 6.3 lays out our considerations on magnetogenesis.

# 6 DISCUSSION

The principal result from the observational analysis is the insignificant dependence of the RM difference of random pairs on source redshift. This requires an additional factor of $(1 + z)^{-\gamma}$ in the RM integral in equation (1) and the best agreement is reached for $\gamma \lesssim 4.5 \pm 0.2$ In the following two sections, we present further motivation for this result.

## 6.1 Mathematical motivation for the redshift evolution of the comoving IGMF

In equation (1), there are several quantities which contain a redshift dependence:

(i) The factor $(1 + z)^{-2}$
(ii) $dl/dz$ depends on the light-travel distance and thus on redshift (e.g. Blasi et al. 1999, equation 2).
(iii) The physical electron number density $n_e(z) = n_{ref}(1 + z)^3$ evolves with redshift like an inverse volume with the cosmological expansion.
(iv) The comoving electron number density $n_{ref}$ also evolves with redshift over the course of structure formation, as Fig. 7 shows.
(v) The physical magnetic field's redshift dependence is given by the flux freezing formula $B = B_0(n_e/n_{ref})^{2/3}$.

Given all of these dependences on redshift, according to the observational data, |ΔRM| appears to be flat with redshift. If all of them together do not yield a flat |ΔRM|, then there must be another quantity that is evolved with redshift, and the only option left is the comoving magnetic field $B_0$. Let us now show with a simple analytical approximation of equation (1) that a redshift evolution like $B_0(z) \equiv B_0(1 + z)^{-\gamma}$ can lead to the desired flat behaviour with $\gamma$ of the order of magnitude that we find in our analysis.

Inserting the IGMF scaling with the electron number density and the expression for $dl/dz$ from Blasi et al. (1999) into equation (1) yields (units suppressed for simplicity and legibility)

$$\text{RM} = 0.812 \int_0^{z_s} c\, B_0(z) \cos\phi\, \frac{n_e^{5/3}(z)}{n_{ref}^{2/3}} \frac{1}{H(z)(1+z)^3}\, dz \quad (5)$$

with $\phi$ being the angle between the magnetic field direction and the directed line element and the comoving IGMF $B_0(z) \equiv B_0(1 + z)^{-\gamma}$. Now we explicitly insert this evolution and we assume a globally homogeneous electron number density, hence $n_e(z) = n_{ref}(1 + z)^3$:

$$\text{RM} = 0.812 \int_0^{z_s} c\, n_{ref}\, B_0 \cos\phi\, \frac{(1+z)^{2-\gamma}}{H(z)}\, dz \quad (6)$$

If we assume, for simplicity, a flat, matter-dominated universe with $H(z) = H_0(1 + z)^{1.5}$ throughout the entire integration range, and a constant magnetic field direction expressed by $\cos\phi$ (e.g. random orientation $\cos\phi = 1/\sqrt{3}$) we can factor all constants[2] out of the

---

[1] Vernstrom et al. (2019) on the other hand find that in the NVSS catalogue, local contributions could potentially be significant since they find a dependence of source polarization fraction $\pi$ on spectral index $\alpha$. Carretti et al. (2022) show that NVSS sources are associated with denser environments than LoTSS DR2 sources, which explains why local contributions could be more significant for NVSS than for LoTSS DR2 sources.

[2] We already noted that $n_{ref}$ is not constant but for the purposes of this analytical approximation we take it as such due to our ignorance of an analytical expression of its redshift dependence.

integral, leaving us with an analytically solvable integral with the solution:

$$\text{RM} = \begin{cases} \frac{0.812}{1.5-\gamma} \frac{c}{H_0} n_{\text{ref}} B_0 \cos\phi \, (1+z_s)^{1.5-\gamma} - C, & \gamma \neq 1.5 \\ 0.812 \frac{c}{H_0} n_{\text{ref}} B_0 \cos\phi \ln(1+z_s), & \gamma = 1.5 \end{cases} \quad (7)$$

with $C = \frac{0.812}{1.5-\gamma} \frac{c}{H_0} n_{\text{ref}} B_0 \cos\phi \approx -0.4 \, \text{rad m}^{-2}$ for fiducial parameter values. The constant $C$ gives an orientation for the expected order of magnitude of the RM which we observe to be well reproduced in, both, the observational data and the Monte Carlo simulations.

As this rough estimate shows, a $\gamma$ of around 1.5 is required to cancel out a potential source redshift dependence of the IGM RM. This result can, of course, only serve for orientation, since there were several approximations:

(i) The calculation took only one source into account. In a random pair picture, this is as if to approximate that the nearer source is at $z = 0$ and the entire $|\Delta\text{RM}|$ of the pair is actually the intergalactic RM of the farther source.

(ii) $n_{\text{ref}}$ was approximated as a constant and taken out of the integration. This approximation shifts the result for $\gamma$ to slightly higher values than if a realistic redshift evolution of the comoving density was taken.

(iii) On the other hand, the result was shifted to significantly lower values by approximating the Universe as matter-dominated along the entire LoS. If the transition to vacuum energy domination at $z \sim 0.5$ is taken into account, the outcome is a higher $\gamma$. This would be a significant correction since around 40 per cent of the simulated sources have redshifts below 0.5.

Consequently, according to this very simple approximative model, if the RM appears to be constant with source redshift, we can deem any result $\gamma \gtrsim 1.5$ plausible.

### 6.2 Motivation for the redshift evolution of the comoving IGMF from cosmological simulations

Here, we make further use of cosmological simulations (Vazza et al. 2017), to extract LoS data of $n_e$ and $B$ up to $z = 2$. We did not generate the $\Delta\text{RM}$ as a function of $z$ directly from the simulations, due to the computational expense of generating the required deep light-cones through the cosmological simulation volumes (e.g. Hackstein, Brüggen & Vazza 2021), the requirement of populating and positioning the radio sources in a realistic manner throughout the simulation volume (e.g. Hodgson et al. 2021), and the need for subgrid modelling of the RM variations in the local environment of the radio sources, all of which was beyond the scope of this work.

We already used the LoS density data to create density distributions used in Section 4.1. Since each LoS also has a comoving magnetic field array as output, we compute the mean comoving magnetic field in 10 redshift bins along each LoS to smooth out small-scale variations (see Fig. 9 for a sample LoS). As we can see from the plot (note the logarithmic scale) the redshift-binned mean magnetic field seems to follow a power-law evolution like $\propto (1+z)^{-\gamma}$. We fit the function $B_m(1+z)^{-\gamma}$ on to the 10 means for each LoS and average the resulting $\gamma$ values over 100 LoSs for each magnetogenesis scenario. The results are displayed in Table 6. As a precaution, we recompute the mean $\gamma$ for each magnetogenesis model by averaging only over the 25 per cent LoS with the highest RM values (the 75th percentile). This is to ensure that the results are not subject to a potential selection bias in the observations due to a lower bound on the detectable RM. The results for $\gamma$ were insensitive to this selection bias.

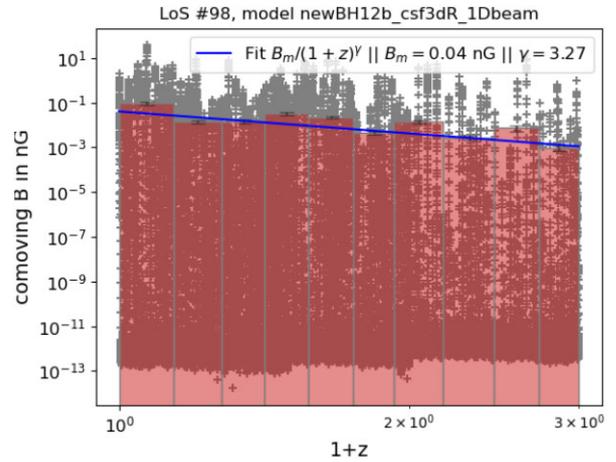

**Figure 9.** Comoving IGMF amplitudes along one particular LoS with an astrophysical magnetogenesis model in cosmological simulations (Vazza et al. 2017). The faded red bars show the mean comoving IGMF amplitude in a given redshift bin and point towards a power-law evolution with redshift. Thus, a fit $\propto (1+z)^{-\gamma}$ (blue line) has been performed to extract the exponent $\gamma$ giving the redshift evolution of the comoving magnetic field. The values presented for each magnetogenesis model in Table 6 result from averaging over the fits on to 100 such LoSs.

**Table 6.** Results for $\gamma$ averaged over 100 LoSs for each magnetogenesis model in cosmological simulations (Vazza et al. 2017).

| Model | $\gamma$ |
| --- | --- |
| Primordial | $-0.26 \pm 0.02$ |
| Dynamo | $4.18 \pm 0.11$ |
| Astrophysical | $2.32 \pm 0.16$ |

The panels in Fig. 10 represent the mean magnetic fields in the 10 redshift bins, averaged over all LoSs for each magnetogenesis model. We see a good confirmation of a power-law behaviour (here in linear scaling). Depending on the model being considered, the comoving magnetic field strength peaks at $z \approx 0$ (dynamo and astrophysical models) or it basically remains constant across the redshift range considered here (primordial model). The latter is the predicted behaviour for the magnetization of cosmic structures close to the mean cosmic density as in the case of filaments of the cosmic web, in which case the evolution is just dictated by adiabatic gas compression. This effect is extremely small at the small overdensity of filaments, thus maintaining the comoving magnetic field close to its assumed primordial value.

On the other hand, the rising trend towards $z = 0$ observed in other two models reflects the gradual increase of the filling factor of significant magnetic fields in filaments, in the two assumed scenarios: either because of the increased level of gas vorticity within growing filaments (dynamo model) or because of the integrated effect of magnetic 'bubbles' blown by the combination of star formation and AGN feedback, which partially percolate within the volume of filaments at lower redshift (astrophysical model). At the qualitative level, the latter trend is the same also as recently reported by Arámburo-García et al. (2021), albeit using different simulations and feedback prescriptions.

A conclusion we can already draw from Fig. 10 and Table 6 is that, while our best-fitting result for $B_0 \lesssim (2.0 \pm 0.2)$ nG is in agreement with the comoving IGMF value for the primordial case of the cosmological simulations (see left-most panel of the

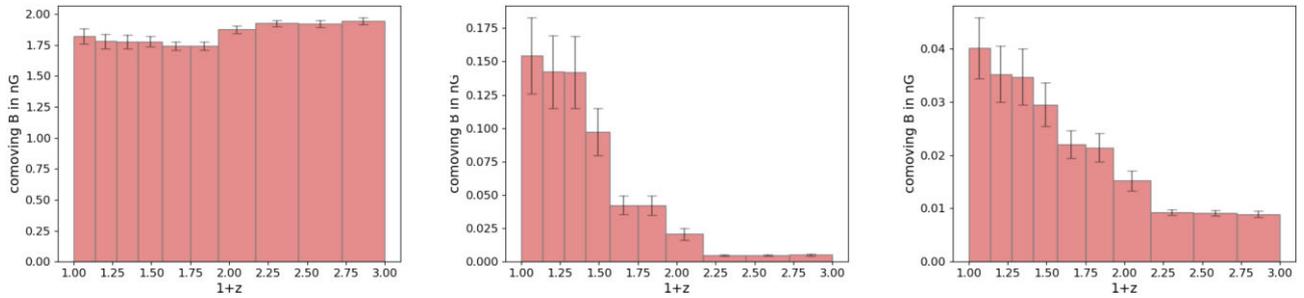

**Figure 10.** Running mean comoving magnetic field amplitudes in 10 redshift bins, averaged over all 100 LoSs for each magnetogenesis model taken from cosmological simulations (Vazza et al. 2017). In all three cases (primordial, dynamo, and astrophysical, shown in this order in the figure).

figure), the evolution exponent's upper limit $\gamma = 4.5 \pm 0.2$ is completely incompatible with the cosmological simulation result in the primordial scenario. To exemplify this: for a $\gamma$ of 4.5, at $z = 2$ the field must already have fallen to around 0.01 nG, which seems to roughly correspond to the result in the dynamo or the astrophysical case of the cosmological simulations. At the same time, our $\gamma \lesssim 4.5 \pm 0.2$ is compatible with the dynamo scenario, whereas our result for $B_0$ is not (Fig. 10), or at least not in the neighbourhood of the upper limit. In the following section, we discuss the physical implications.

### 6.3 Implications for magnetogenesis

#### 6.3.1 Previous work in context

The process for the origin of IGMFs is poorly constrained. Recently, SOS20 found a strong dynamo amplification of a weak primordial magnetic field to be disfavoured by LOFAR data. They compared the structure functions $\langle \Delta RM(\Delta \theta)^2 \rangle$ of cosmological simulations, for the same three magnetogenesis scenarios considered here, to the structure function underlying the observational data (see SOS20's fig. 8). In the dynamo scenario considered in SOS20, which is amplification by the solenoidal turbulence developed in large-scale structures, the structure function is irreconcilably high with respect to the data, even though it is important to note that there are very many dynamo amplification mechanisms (Donnert et al. 2018) and this tension does not immediately dismiss all other dynamo mechanisms.

Meanwhile, primordial magnetogenesis has received support by the reported direct detection by Vernstrom et al. (2021) of a surprisingly high synchrotron emission flux coming from filaments, although note that Hodgson et al. (2021) have tried to reproduce this excess emission using the FIGARO simulations but could not. This discrepancy between the empirical and the simulation results remains unresolved for now. Nevertheless, from the measured inverse Compton and synchrotron flux densities under the assumption of equipartition, Vernstrom et al. (2021) estimate the magnetic field required to produce this radiation to 30–60 nG. Note that they claim this to be in conflict with upper bounds of a few nano-Gauss derived in sources like in O'Sullivan et al. (2020), although see also Brown et al. (2017) and Hackstein et al. (2016). This is not necessarily the case because the values reported in these papers essentially reflect an expectation value of the IGMF across the entire IGM. Since the IGM stands for both filaments and voids, the latter of which shifts this expectation towards lower values, it is not surprising that the reported results for the magnetic field are different. Inserting a number density value of $5 \times 10^{-5}$ cm$^{-3}$ (see e.g. Gheller et al. 2015) typical of the densest regions expected in filaments into the IGMF density scaling $B = B_0(n_e/n_{ref})^{2/3}$ underlying the SOS20 model yields, for a typical $B_0$ of 1 nG, a magnetic field $B$ of 40 nG, which fits within the value range quoted in Vernstrom et al. (2021).

Until the exact population of objects responsible for the excess in Vernstrom et al. (2021) is known for sure, it is difficult to make definitive conclusions. Nevertheless, the detection of a field strength of 30–60 nG in filaments would be highly indicative of a non-vanishing primordial contribution to the IGMF because the dynamo amplification and the astrophysical scenarios on their own are unable to produce magnetic fields of an amplitude this high in filaments as opposed to in the interiors or immediate surroundings of galaxies and galaxy clusters (see e.g. fig. 3 in Vazza et al. 2017). Even though the above detected amplitude could imply a non-vanishing primordial seed field, a primordial model alone (a field strength of order nG at high redshift that is modified only by structure formation) without further amplification by a dynamo or by astrophysical sources is in most cases also unable to produce field strengths of the above amplitude, unless the primordial seed field is at the very top of the present upper limits (Vazza et al. 2015, 2017). Carretti et al. (2022) follow this line of thought. They opt for a single-source RM approach to the LoTSS DR2 144 MHz data and conclude that the LoTSS DR2 RM data correspond to low-density environments and find a flat dependence of the observed RM with $z$. They estimate the mean magnetic field strength in cosmic filaments to be ∼30 nG, which is compatible both with our crude estimate above of 40 nG and with Vernstrom et al. (2021)'s result, and argue that this value points towards a combined primordial and astrophysical magnetogenesis by comparison with cosmological simulations (e.g. Vazza et al. 2015, 2017; Arámburo-García et al. 2021).

Let it be noted again that the simulations by Vazza et al. (2017) for the primordial, astrophysical, and dynamo cases are just three specific realizations of those magnetogenesis mechanisms. For example, there could be primordial models with uniform seed fields lower than ∼1 nG, or with tangled, non-uniform seed fields, as in Vazza et al. (2020), or with a power-law distribution of the initial magnetic field fluctuations, as explored in Vazza et al. (2021), or astrophysical models with higher magnetic energy injection than the one taken by Vazza et al. (2017). Finally, there can also be 'combined' models, such as including both a primordial seed field and a consecutive astrophysical magnetic energy injection, as put forth by Carretti et al. (2022).

#### 6.3.2 This work

Although we merely parametrized the evolution of the comoving IGMF, our work complements the larger efforts of simulating the RM sky in full cosmological simulations by adding real data and comparing their features to simulation results.

Our simulations and analysis show that primordial magnetogenesis with uniform seed fields $B_0 \gtrsim 0.01$ nG at $z > 2$ are disfavoured if we take $B_0$ and $\gamma$ to be at the upper limits we derive: While $B_0$ (the comoving field strength at $z = 0$) of $\lesssim (2.0 \pm 0.2)$ nG is compatible with Vazza et al. (2017)'s cosmological simulation realization of a primordial magnetogenesis model (see again Fig. 10), with a $\gamma$ of 4.5 this value will have dropped to ∼0.01 nG by $z = 2$, which puts new important limits to any model of primordial magnetogenesis predicting fields correlated on cosmological scales (like in the case of inflationary mechanisms).

We cannot constrain the magnetogenesis scenario further than this, although the maximum likelihood method for model selection exhibits a preference for *dyn* models. First of all, the different magnetogenesis scenarios are explicitly represented in our simulation model solely by the density values as taken from the cosmological simulations. The top five best-fitting models are not exclusively with densities from the *dyn* model, instead all three density distributions are represented in the top five. We furthermore argued that the absence of local-to-source effects in our simulations is probably leading to a drop in the absolute RM difference towards low $z_<$ which is not present in the observational data. It could be that including a small local contribution into our simulation model 'fixes' the flatness behaviour of all of our models and thus makes *dyn* models even more preferable, but until local contributions are implemented into our simulations it will remain unclear what change this would bring about in the preference of a given magnetogenesis scenario. Therefore, we agnostically refrain from claiming that our analysis is capable of discriminating between different scenarios of magnetogenesis other than disfavouring uniform seed fields of order nG by finding a preference for a steep redshift evolution of the magnetic field. On the contrary, our findings could be implying that, purely based on the density distributions, it is difficult to distinguish between the different magnetogenesis models in terms of the resulting comoving IGMF value today and its redshift evolution exponent.

## 7 CONCLUSIONS

We presented an analysis of the LoTSS DR2 RM data with redshift information using a random pair approach, with physical pairs serving as a control sample. We find a median absolute RM difference for random pairs $|\Delta RM|_{med,rp} = (1.79 \pm 0.09)$ rad m$^{-2}$ after a correction for the Galactic contribution, and a median $|\Delta RM|$ excess of random over physical pairs of $(1.65 \pm 0.10)$ rad m$^{-2}$. We cannot reject the hypothesis of $|\Delta RM|$ being flat with respect both to the redshift difference of the pair and to the redshift of the nearer source. The physical pair control sample has $|\Delta RM|_{med,pp} = (0.70 \pm 0.08)$ rad m$^{-2}$, which indicates that the IGM is not the only factor contributing to the RM difference for RPs, but contributions from local environments play a smaller but non-negligible role as well. This is in agreement with results by Carretti et al. (2022), who used the same data but adopted the single-source RM rather than the paired-source approach.

By comparing the results from the observational data analysis and our Monte Carlo simulations, we confirm results from previous work (Planck Collaboration XIX 2016b; O'Sullivan et al. 2020) that derived upper limits of order 1 nG for the volume-filling comoving IGMF ($B_0$) at $z = 0$. We find a best-fitting value of $B_0 \lesssim (2.0 \pm 0.2)$ nG for a coherence length factor $l_0 \lesssim 0.10 \pm 0.07$, whose inverse stands for the number of times the magnetic field changes orientation along the physical (cosmologically redshift-evolved) Jeans length. The results are quoted as upper limits due to since we model only the intergalactic contribution on Mpc-scales and larger and do not model local contributions, which turn out to play an important role in ensuring a flat behaviour of low $|\Delta RM|$ at low source redshifts. Furthermore, compensation effects between the parameters tend to cause a preference for higher values.

Most notably, we find a redshift evolution of the comoving IGMF that follows $\propto (1 + z)^{-\gamma}$, with $\gamma \lesssim 4.5 \pm 0.2$, which is compatible with the value $(4.18 \pm 0.11)$ that we extract from LoSs from Vazza et al. (2017)'s cosmological simulations with an underlying dynamo magnetogenesis model.

By comparing results with cosmological simulations by Vazza et al. (2017), we showed that uniform primordial seed field strengths of order nG are disfavoured by the data. Due to the absence of local-to-source effects, our model does not further distinguish between the different magnetogenesis scenarios just based on the density distributions they produce.

## ACKNOWLEDGEMENTS

VPP acknowledges funding by DAAD-RISE (Deutscher Akademischer Austauschdienst - Research Internships in Science and Engineering) for funding of the project. MB acknowledges funding by the Deutsche Forschungsgemeinschaft (DFG, German Research Foundation) under Germany's Excellence Strategy - EXC 2121 'Quantum Universe' – 390833306. Simulations of this work used ENZO code (http://enzo-project.org), the product of a collaborative effort of scientists at many universities and national laboratories. Our simulations were run on the Piz Daint supercomputer at CSCS-ETH (Lugano). FV acknowledges financial support from the Horizon 2020 program under the ERC Starting Grant 'MAGCOW', no. 714196. This work has been conducted within the LOFAR Magnetism and Surveys Key Science Projects (MKSP; https://lofar-mksp.org/, SKSP; https://lofar-surveys.org/). This research made use of AS-TROPY, a community-developed core PYTHON package for astronomy (Astropy Collaboration 2013) hosted at http://www.astropy.org/, of MATPLOTLIB (Hunter 2007), and of TOPCAT, an interactive graphical viewer and editor for tabular data (Taylor 2005). This research has made use of ALADIN SKY ATLAS developed at CDS, Strasbourg Observatory, France (Bonnarel et al. 2000). LOFAR (van Haarlem et al. 2013) is the Low Frequency Array designed and constructed by ASTRON. It has observing, data processing, and data storage facilities in several countries, which are owned by various parties (each with their own funding sources), and that are collectively operated by the ILT foundation under a joint scientific policy. The ILT resources have benefited from the following recent major funding sources: CNRS-INSU, Observatoire de Paris and Université d'Orléans, France; BMBF, MIWF-NRW, MPG, Germany; Science Foundation Ireland (SFI), Department of Business, Enterprise and Innovation (DBEI), Ireland; NWO, The Netherlands; The Science and Technology Facilities Council, UK; Ministry of Science and Higher Education, Poland; The Istituto Nazionale di Astrofisica (INAF), Italy. This research made use of the Dutch national e-infrastructure with support of the SURF Cooperative (e-infra 180169) and the LOFAR e-infra group. The Jülich LOFAR Long Term Archive and the German LOFAR network are both coordinated and operated by Jülich Supercomputing Centre (JSC), and computing resources on the supercomputer JUWELS at JSC were provided by the Gauss Centre for Supercomputing e.V. (grant CHTB00) through the John von Neumann Institute for Computing (NIC). This research made use of the University of Hertfordshire high-performance computing facility and the LOFAR-UK computing facility located at the University of Hertfordshire and supported by STFC [ST/P000096/1], and of the Italian LOFAR IT computing infrastructure supported and operated by INAF, and by the Physics Department of Turin university (under an agreement with Consorzio Interuniversitario per la Fisica Spaziale) at the C3S Supercomputing Centre, Italy. The authors thank the referee for constructive comments which significantly improved the quality of the paper.

## DATA AVAILABILITY

The LoTSS-DR2 data used in this work will be available in O'Sullivan et al. (in preparation). The pairs catalogue made from these data is available on VizieR. The Monte Carlo simulation code is available upon request, and the R code used for correlated draws from the cosmological density distributions is available here: https://gist.github.com/nicebread/4045717. Examples of the data set of ENZO simulations analyzed in this work can be found at this URL: https://cosmosimfrazza.myfreesites.net/scenarios-for-magnetogenesis.

## APPENDIX A: PHYSICAL EFFECTS BEHIND THE DIFFERENT DENSITY DISTRIBUTIONS

It is important to note, once again, that the three magnetogenesis scenarios are represented in our model only through the density drawing; they are in no way modelled explicitly. Thus, differences in our simulation results from models with the only difference being the density drawing directly reflect the differences in the density distributions that the three magnetogenesis scenarios as taken from Vazza et al. (2017) produce.

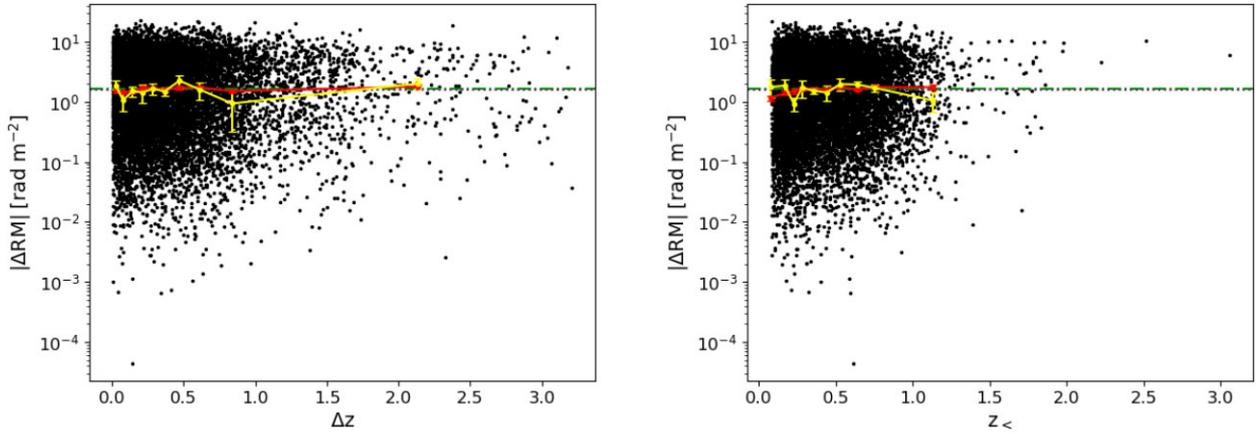

**Figure A1.** Scatter of the absolute RM difference of random pairs versus redshift difference (left-hand panel) and versus the lower source redshift (right-hand panel) for the second-best-fitting result | prim | 1.0 | 10 | 4.5 | The representation is identical to Fig. 8.

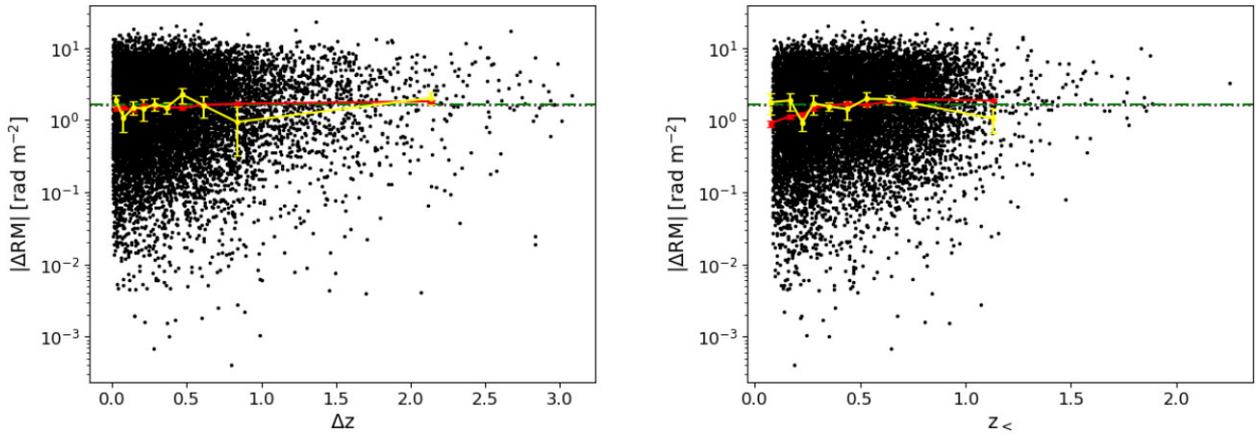

**Figure A2.** Scatter of the absolute RM difference of random pairs versus redshift difference (left-hand panel) and versus the lower source redshift (right-hand panel) for the fifth-best-fitting result | astro | 2.0 | 10 | 3.0 | The representation is identical to Fig. 8.

Clearly, while the gas density distributions on large scales look very similar in all models (Fig. 7), the different physical implementations in each run play a detectable role in affecting the density of ionized gas, and hence all RM related observables, along long simulated LoSs. First, only in the simulation with the astrophysical scenario radiative cooling, star formation, and supermassive black holes are included, as well as their energy feedback on the surrounding gas. This can lead both to more gas clumping compared to the other non-radiative models, as well as to a reduced ionized gas density along the LoS. Two factors are responsible for the latter.

First, ionized gas above a given high density threshold is removed from within haloes, due to its assumed instantaneous recycling into stellar population particles and/or black hole particles (Kim et al. 2011). Secondly, feedback from star formation and super massive black holes have an integrated effect also on the scale of filaments, whose internal density structures gets smoothed even on scales of several Megaparsecs from the sources of feedback (e.g. Gheller & Vazza 2019), albeit the precise amplitude of this effect can vary with feedback prescriptions and is still uncertain (e.g. Galárraga-Espinosa et al. 2020).

In addition to the above effects, the very different level of magnetization of voids assumed in the different scenarios can further induce density differences. In particular, while the dilution of magnetic fields ejected from galaxies in the astrophysical scenario typically makes the magnetic field pressure largely subdominant compared to gas pressure everywhere, the much larger magnetic fields that are reached in the other two models makes magnetic pressure often larger than gas pressure in the cold, low density distribution of the IGM (Banfi, Vazza & Wittor 2020). This is an additional source of density variations in the primordial and dynamo models investigated here, on spatial scales which depend on the correlation scales of the magnetic fields in voids and filaments ($\leq 10$ Mpc), and emerge in the integration along long LoSs.

This paper has been typeset from a TEX/LATEX file prepared by the author.